\documentclass[twoside,twocolumn,aps,prl]{revtex4-1}
\usepackage[latin9]{inputenc}
\usepackage{graphicx}
\usepackage{amssymb}
\usepackage{amsmath}
\usepackage{amsfonts}
\usepackage{epsfig}
\usepackage{wrapfig}
\usepackage{bm}
\usepackage{hyperref}
\usepackage[margin=1in,nofoot]{geometry}

\newcommand {\be}{\begin{equation}}
\newcommand {\ee}{\end{equation}}
\newcommand {\bea}{\begin{eqnarray}}
\newcommand {\eea}{\end{eqnarray}}

\usepackage[margin=1in,nofoot]{geometry}

\begin{document}

\title{Dissipative Universal  Lindbladian Simulation}

\author{Paolo Zanardi, Jeffrey Marshall, Lorenzo Campos Venuti}
\affiliation{Department of Physics and Astronomy, and Center for Quantum Information
Science \& Technology, University of Southern California, Los Angeles,
CA 90089-0484}
\begin{abstract}
It is by now well understood that quantum dissipative processes can be harnessed and turned into a resource for quantum-information processing  tasks.
In this paper we demonstrate yet another way in which this is true by providing a  dissipation-assisted protocol for the simulation of general
Markovian dynamics. More precisely, we show how a suitable coherent coupling of a quantum system to a set of Markovian dissipating qubits allows
 one to enact an effective Liouvillian generator of any Lindbladian form.  This effective dynamical  generator arises from high-order virtual-dissipative processes
and governs the system dynamics exactly in the limit of infinitely fast dissipation. Applications to the simulation of collective decoherence are discussed
as an illustration.

\end{abstract}
\maketitle

Quantum  decoherence and dissipation have been regarded until recently purely detrimental to the aim of quantum information processing (QIP) \cite{Unruh:1995fk,Aharonov:96a}.
Interactions with the environment in fact inevitably lead to  entanglement between the quantum computing system and  uncontrollable degrees of freedom.
This unwanted entanglement in turn results in a system subdynamics that is in general incoherent and irreversible: unitarity is quickly lost and with it the quantum information processing  advantages e.g., computational speed-ups, one was seeking for. This state of affairs triggered  a spectacular theoretical effort that led to the discovery of a host of techniques to tame decoherence \cite{Lidar-Brun:book} as quantum error correction \cite{Shor:1995fb,Gottesman:1996fk}, decoherence-free subspaces \cite{Zanardi:97c,Lidar:1998fk,Kielpinski:01}, noiseless subsystems \cite{Knill:2000dq,Zanardi:99d,Kempe:2001uq,Zanardi:2003c,Viola:2001sp} and holonomic quantum computation \cite{HQC, Duan:2001ff}.

It is therefore a conceptually remarkable shift the recent realization that by reservoir-engineering dissipation can be harnessed and turned into a useful practical resource for QIP 
(see \cite{beige} for an early pioneering insight). For example one can dissipatively achieve quantum state preparation \cite{Kraus-prep,kastoryano2011dissipative}, quantum simulation \cite{barreiro2011open}, holonomic quantum computation \cite{Yale}   and even universal computation \cite{verstraete2009quantum}. Simulation of highly non-trivial properties of matter as topological order \cite{dissi-top} and non-abelian synthetic gauge fields \cite{Stannigel:2014rw} can also be accomplished by dissipative means.
Finally, all forms of QIP that encode information in the ground state of a time-dependent Hamiltonian, e.g., open system adiabatic quantum computation and quantum annealing, also benefit from dissipation and relaxation to negate thermally driven errors \cite{childs_robustness_2001,PhysRevLett.95.250503,PAL:13}.

In particular in \cite{zanardi-dissipation-2014} it has been shown that quantum information can be encoded in the set of steady states (SSS) of a sufficiently symmetric strongly dissipative system and manipulated coherently by an effective dissipation-projected Hamiltonian. The latter is  of geometric nature and  is robust against some types of Hamiltonian and dissipative perturbations \cite{zanardi-emerging-2014}. 
The key idea of Ref. \cite{zanardi-dissipation-2014} is a simple one: once the system is prepared in the SSS the fast dissipative processes adiabatically decouple non steady-states away while at the same time strongly renormalize the system Hamiltonian in such a way that the SSS remains invariant under this projected dynamics. This phenomenon can be thought of as a sort of 
{environment-induced} quantum Zeno effect \cite{Facchi:PRL02,PhysRevLett.108.080501} at the superoperator space level \cite{zanardi-emerging-2014}. 

\begin{figure}[h]
\vspace{-5mm}
\begin{centering}
\includegraphics[width=8cm]{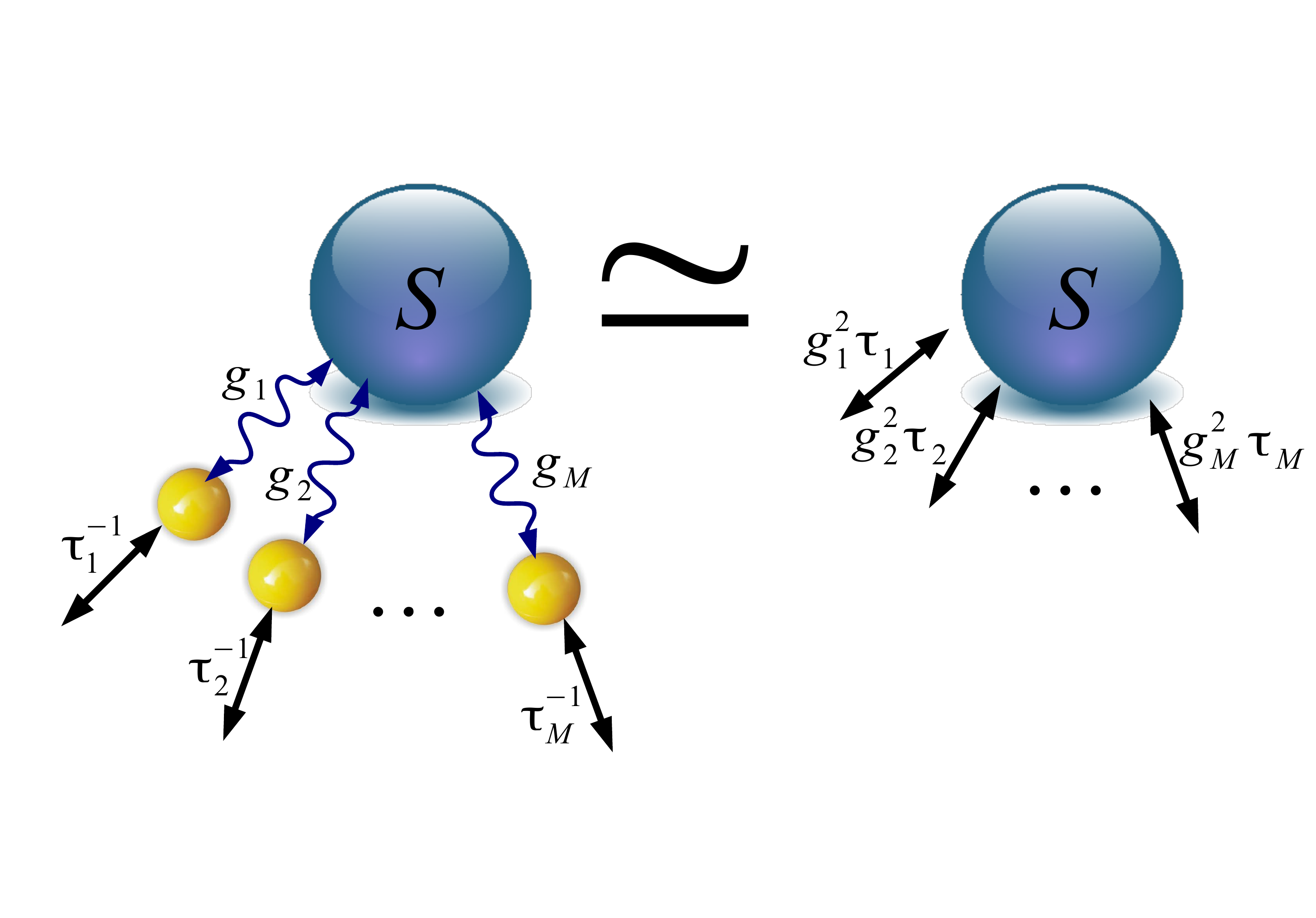}
\end{centering}
\vspace{-13mm}
\caption{A quantum system $S$ (blue ball) is coupled with coupling strengths $g_i$ to $M$  qubits (yellow balls). Each of these qubits is subject to amplitude damping with rates $\tau_i^{-1}$.
Proposition 2 shows that in the limit of small $\tau_i$ the qubits can be adiabatically decoupled and the effective dynamics of $S$ is described by $M$ Lindblad operators of strength $g_i^2 \tau_i$
\label{2nd-Order}}
\end{figure}

In this paper we extend the ideas of \cite{zanardi-dissipation-2014} to higher order. In the case in which the dissipation-projected Hamiltonian is vanishing,
higher order virtual dissipative processes give rise, in a suitable limit, to an effective Liouvillian generator that leaves the SSS invariant. However, at variance with the case studied in \cite{zanardi-dissipation-2014} this effective generator is no longer Hamiltonian: a slow irreversible process unfolds within the SSS. We will show how this mechanism can be exploited to the end
of the simulation of {\em{any}} Markovian dynamics. More precisely, we will show that by suitably coupling a quantum system to a structured reservoir comprising multiple qubits undergoing fast amplitude damping one can implement an effective Liouvillian generator in general Lindblad form \cite{Lindblad-paper}. We will illustrate our results by analyzing the dissipative simulation of qubits subject to collective decoherence.


{\em{Preliminaries.--}}
Let ${\cal H,\,[\mathrm{dim}({\cal H})<\infty]}$ 
denote the Hilbert space of the system and ${\mathrm{L}}({\cal H})$
the algebra of linear operators over it. A time-independent Liouvillian
super-operator ${\cal L}_{0}$ acting on L$({\cal H})$ is given.
The SSS  of ${\cal L}_{0}$ consists of all the quantum states $\rho$ ($\rho\in {\mathrm{L}}({\cal H}),\,\rho\ge 0,\,{\mathrm{Tr}}\,\rho=1$)
contained in the kernel ${\mathrm{Ker}}\,{\cal L}_{0}:=\{X\,/\,{\cal L}_0(X)=0\} $ of ${\cal L}_0$. 
%
We shall denote by ${\cal P}_{0}={\cal P}_0^2$ (${\cal Q}_{0}:=1-{\cal P}_{0}$)
the spectral projection over Ker$\,{\cal L}_{0}$ (the complementary
subspace of Ker$\,{\cal L}_{0}$.)
As in \cite{zanardi-dissipation-2014} the Liouvillian ${\cal L}_{0}$ is also assumed to be such that: 
\textbf{{a)}}
$e^{t{\cal L}_{0}},\,(t\ge0)$ defines
a semi-group of trace-preserving positive maps with $\|e^{t{\cal L}_{0}}\|\le1$, 
\textbf{{b)}} The non-zero eigenvalues $\lambda_{h},\,(h>0)$
of ${\cal L}_{0}$ have negative real parts, i.e., the SSS is attractive.
In this case ${\cal P}_{0}=\lim_{t\to\infty}e^{t{\cal L}_{0}}$ and 
${\cal P}_{0}\,{\cal L}_{0}={\cal L}_{0}\,{\cal P}_{0}=0.$
We also denote by ${\cal S}:=-\lim_{z\to 0} (z-{\cal L}_0)^{-1} {\cal Q}_{0}$  the  reduced resolvent of ${\cal L}_0$
at ($z=0$)   and by $\tau_R:=\|{\cal S}\|.$ 
The latter provides a natural time-scale associated with the relaxation processes described by ${\cal L}_0.$
The energy scale  $\tau_R^{-1}$ is of the order of the dissipative gap of ${\cal L}_0$ i.e., the smallest modulus of a non-zero eigenvalue of ${\cal L}_0.$
The dimensionless (and normalized) resolvent is defined by $\tilde{S}:= \tau_R^{-1} {\cal S}.$
%
%
%
%
We now add an Hamiltonian term ${\cal K}:=-i[K,\,\bullet]$
where $K=K^{\dagger}$ such that ${\cal L}_{T}={\cal L}_0 +{\cal K}$.  We set ${\cal K}= (\tau_R T )^{-1/2} \tilde{\cal K},$ in such a way that $\tilde{\cal K}$ is dimensionless and $\|\tilde{\cal K}\|=O(1).$ The time-scale $T$
is our scaling parameter and has to be thought of as large or even infinite in the spirit of the adiabatic theorem.
 %
 We first establish 
%

\noindent \textbf{Proposition 1}: If ${\cal P}_{0}{\cal K}{\cal P}_{0}=0,$ then for sufficiently large $T$ one has that
\begin{equation}
\sup_{t\in[0,\,\theta T]}\|(   e^{t {\cal L}_T}-e^{\frac{t}{T}\tilde{{\cal L}}_{{\textrm{\textrm{eff}}}}}){\cal P}_{0}\|\le C_\theta \sqrt{\frac{\tau_R}{T}} \label{2nd-projection-th}
\end{equation}
 where  $\tilde{{\cal L}}_{{\textrm{eff}}}:= -{\cal P}_{0}{\tilde{\cal K}}\tilde{\cal S}{\tilde{\cal K}}{\cal P}_{0},$
$ {\cal L}_T:= {\cal L}_0 +\frac{1}{\sqrt{\tau_R T}} \tilde{{\cal{K}}},$  $\theta=O(1)>0$  and $C_\theta=O(1)$ depends on $\theta$ and ${\cal L}_0.$

{\em{Proof.--}} 
Is provided in the  Appendix. $\hfill\Box$ 

This  result provides the starting point of this paper.
In particular from Eq.~(\ref{2nd-projection-th}) it follows  that
$\lim_{T\to\infty}\|(e^{T\theta  {\cal L}_T}- e^{\theta\tilde{\cal{L}}_{\textrm{eff}}}){\cal P}_0\|=0.$
In words: if the system is prepared at time $t=0$ inside the SSS and then evolves for finite fraction $\theta$ of $T$, in the large $T$ limit the time-evolution leaves the SSS invariant and it is governed by the effective (dimensionless) generator  $\tilde{{\cal L}}_{{\textrm{eff}}}$.
[It is also sometimes convenient to introduce the effective dimensionful generator  ${{\cal L}}_{{\textrm{eff}}}:= -{\cal P}_{0}{{\cal K}}{\cal S}{{\cal K}}{\cal P}_{0}$
whose norm is $\|{{\cal L}}_{{\textrm{eff}}}\|=O(\|K\|^2\tau_R).$  In terms of ${\cal L}_{\textrm{eff}} $ the second term in the norm of Eq. (\ref{2nd-projection-th}) reads $e^{t {\cal L}_{\textrm{eff}} }.$]

{\em{Remarks:}}
{\bf{0)}} The stronger the dissipation outside the SSS 
i.e., the shorter $\tau_R,$ the {\em{weaker}} the effective one inside
{\bf{i)}} Since, by construction $\|\tilde{\cal{L}}_{\textrm{eff}}\|=O(1),$ the (dimensionless) action associated to the effective propagator  $e^{\theta\tilde{\cal{L}}_{\textrm{eff}}}$ is $O(\theta)$ for $T\to\infty.$  {\bf{ii)}}
The RHS of Eq. (\ref{2nd-projection-th}) represents an error bound, if we fix it at $\epsilon\ll1,$ we see that one needs that $T\ge\epsilon^{-2}\, {C_\theta\tau_R}.$
{\bf{iii)}}  If ${\cal K}\mapsto {\cal K}+T^{-1} \tilde{\cal K}_1$  where $\|\tilde{\cal K}_1\|=O(1)$ and ${\cal P}_0 {\cal K}_1{\cal P}_0\neq 0$
then (\ref{2nd-projection-th}) holds with $\tilde{\cal L}_{\textrm{eff}}\mapsto \tilde{\cal L}_{\textrm{eff}}+ {\cal P}_0 \tilde{\cal K}_1{\cal P}_0$ and a different constant $C_\theta=O(1)$
\cite{add-ham}


The effective Liouvillian generator  $\tilde{\cal{L}}_{\textrm{eff}}$ is clearly reminiscent of the second-order effective Hamiltonians routinely used e.g., in quantum optics, and obtained by some sort of adiabatic decoupling technique \cite{Gardiner-Zoller}. However, this dynamics, at variance with that case as well as with the  situation  considered in \cite{zanardi-dissipation-2014} is not unitary but of general Liouvillian type. The key point is that this effective non-unitary dynamics depends
on ${\cal K}$ and on its non-trivial interplay with the bare dissipation generated by ${\cal L}_0.$
This opens up the possibility of  using it to {{engineer dissipative systems}} with a desired Liouvillian  generator.


{\em{Universal Lindbladian simulation.--}}
Let us consider a system $S$ coupled to a system $B$ via the general Hamiltonian
\begin{equation}
K=\sum_{i=1}^M \, L_{i} \otimes B_i,
\label{K-simul}
\end{equation}
where the tensor ordering follows that of the total Hilbert space, $\cal H = \cal H_S \otimes \cal H_B$ and, without loss of generality, we assume $B_i^\dagger=B_i,\,(i=1,\ldots, M)$. We also assume that the dissipative term is of the form ${\cal L}_0={\mathbf{1}}_S\otimes {\cal L}_B$, such that ${\cal L}_B (\rho_0)=0$ where $\rho_0$ is by assumption the unique steady state of ${\cal L}_B$. The SSS of ${\cal L}_0$ is given by all the states of the form $\rho\otimes \rho_0$ and it is  isomorphic to the full-state space of $S.$
In this case one has  ${\cal P}_0 (X) = {\mathrm{Tr}}_B(X) \otimes {\rho}_0$ and ${\cal P}_0 {\cal K}{\cal P}_0(\bullet)=-i[K_{\mathrm{eff}},\bullet]$
with $K_{\mathrm{eff}}=\mathrm{Tr}_B\left( K\rho_0\right)\otimes{\mathbf{1}}_B$\cite{zanardi-dissipation-2014}. Let ${\cal S}_B$ be the  the projected resolvent of ${\cal L}_B$ at $z=0.$ 

{\bf{Proposition 2}:} If $K_{\mathrm{eff}}=0$ then  ${\cal L}_{\textrm{eff}}={\cal L_{\textrm{eff}}^{(S)}}\otimes {\mathbf{1}}_B,$ 
\begin{equation} 
{\cal L}^{(S)}_{\textrm{eff}}(\rho) = -i[H_{\textrm{eff}},\rho] + \sum_{i,j=1}^M2\Gamma_{ij}(L_i\rho L_j - \frac{1}{2}\{L_jL_i\,,\rho \})
\label{2nd-order-general-2}
\end{equation}
where $\Gamma :=  (\Gamma^{(A)}+\Gamma^{(A)\dagger})/2$,  $H_{\textrm{eff}} = \frac{1}{2i}\sum_{i,j=1}^M(\Gamma^{(A)} - \Gamma^{(A)\dagger})_{i,j}L_jL_i$
and  $\Gamma_{ij}^{(A)} = -{\mathrm{Tr}}\left( {\cal S}_B(B_i \rho_0)B_j\right).$ 

{\em{Proof.--}} Is provided in the Appendix. $\hfill\Box$

Notice that  $H_{\textrm{eff}}^\dagger=H_{\textrm{eff}}$ and that Eq.~(\ref{2nd-order-general-2}) describes a truly Lindbladian dynamics iff $\Gamma\ge 0.$
Our main result now follows as a particular case of Proposition 2 above. Let us consider   a $d$-dimensional system $S$ coupled to a system $B$ comprising $M$  qubits, by the Hamiltonian
$K=\sum_{i=1}^M g_i\,( L_{i}^\dagger \otimes \sigma_i^-+{\mathrm{h.c.}}),$
where the $L_i$'s are given operators acting on the system state-space only. 
 Let us also suppose ${\cal L}_B=\sum_{i=1}^M{\cal L}_i$ where
each of the $M$ qubits independently dissipates according to the local Liouvillian 
\begin{equation}
{\cal L}_i(\rho)=\tau_i^{-1} (\sigma_i^-\rho\sigma_i^+ -\frac{1}{2}\{\sigma^+_i\sigma^-_i,\,\rho\}).
\label{qubit-damping}
\end{equation}
The unique steady state of ${\cal L}_B$ is $\rho_0=|0\rangle\langle 0|^{\otimes\,M}$ and since ${\mathrm{Tr}}( \sigma_i^{\pm}  \rho_0)=0\,(\forall i$) one has 
$K_{\mathrm{eff}}=0.$ 

{\bf{Proposition 3: }} 
${\cal L}_{\textrm{eff}}={\cal L_{\textrm{eff}}^{(S)}}\otimes {\mathbf{1}}_B$ where
\begin{equation}
{\cal L}^{(S)}_{\textrm{eff}}(\rho)=4\sum_{i=1}^M {g_i^2\tau_i}( L_i\rho L_i^\dagger-\frac{1}{2}\{L_i^\dagger L_i,\,\rho\}).
\label{L-simul}
\end{equation}
{\em{Proof.--}} 
To obtain Eq. (\ref{L-simul}) from Eq.~(\ref{2nd-order-general-2}), re-write the $B_i,L_i$  in (\ref{K-simul}) such that $K=\sum_{i=1}^M g_i\,( L_{i}^\dagger \otimes \sigma_i^-+{\mathrm{h.c.}})$. Remembering that $\rho_0 = |0\rangle\langle 0|^{\otimes\,M}$, and ${\cal L}_B$ as in Eq.~\eqref{qubit-damping}, we recover Eq.~\eqref{L-simul} as required  by direct evaluation  of the matrix $\Gamma^{(A)}$ in Prop.~2 $\hfill\Box$ \\

Notice now that, in view of remark {\bf{iii)}} after Prop.~2, one can add any Hamiltonian ${\cal K}_1=T^{-1} \tilde{{\cal K}}_1$ [$\|\tilde{\cal K}_1\|=O(1)$]
acting on the system $S$ only ($\Rightarrow{\cal P}_0 {\cal K}{\cal P}_0= {\cal P}_0 {\cal K}_1 {\cal P}_0$). This will result in $\tilde{\cal L}_{\textrm{eff}}\mapsto \tilde{\cal L}_{\textrm{eff}}+\tilde{\cal K}_1.$ Therefore we see that Prop.~2 shows that in principle {\em{any}} Liouvillian in the Lindblad form \cite{Lindblad-paper} i.e., the most general generator
of semi-groups of Markovian CP maps, can be obtained given the availability of $M$ auxiliary qubits (one for each Lindblad operator) 
subject to an amplitude damping channel  {\em{and }} the ability to enact the Hamiltonian $K$. Dissipation turns into a resource that allows one to {\em{simulate}} a general Lindbladian evolution. 

 We would like to make a couple of remarks: 
{\em{1)}}  One might think of obtaining the Lindbladian dynamics Eq.~(\ref{L-simul}) {\em{directly}} coupling the system $S$ to some reservoir with an interaction Hamiltonian of type (\ref{K-simul})  and then using the standard Born Markov approximation \cite{Gardiner-Zoller}. The point is that the latter involves {\em{uncontrolled}} approximations (Markov) whereas Eq.~(\ref{2nd-projection-th}) has a uniform and {\em{controlled}} error \ $O(\sqrt{\tau_R/T}).$  This means that the effective dynamics of $S$ becomes {\em{exactly}} Lindbladian, with generator (\ref{L-simul}), for $T\to\infty.$ Of course this is true as long as the auxiliary qubits are {\em{exactly}} described by the Lindbladian in Eq.~(\ref{qubit-damping}) i.e., their genuine Markovianity is a key resource in our universal simulation protocol along with the ability of switching on the the Hamiltonian in Eq.~(\ref{K-simul}).
{\em{2)}} In view of physical applications, we stress  that the effective dynamics in Eq.~(\ref{L-simul}) still holds if the $M$ qubits are replaced by $M$ bosonic modes subject
to amplitude damping i.e., the $\sigma_i^-$ in Eq.~(\ref{qubit-damping}) are replaced by annihilation operators $a_i$.

In the next section we will discuss, for the sake of illustration of our general results, the simulation of different types of collective decoherence
when $S$ is itself a set of multiple qubits.



\begin{figure}
\begin{centering}
\includegraphics[width=8cm]{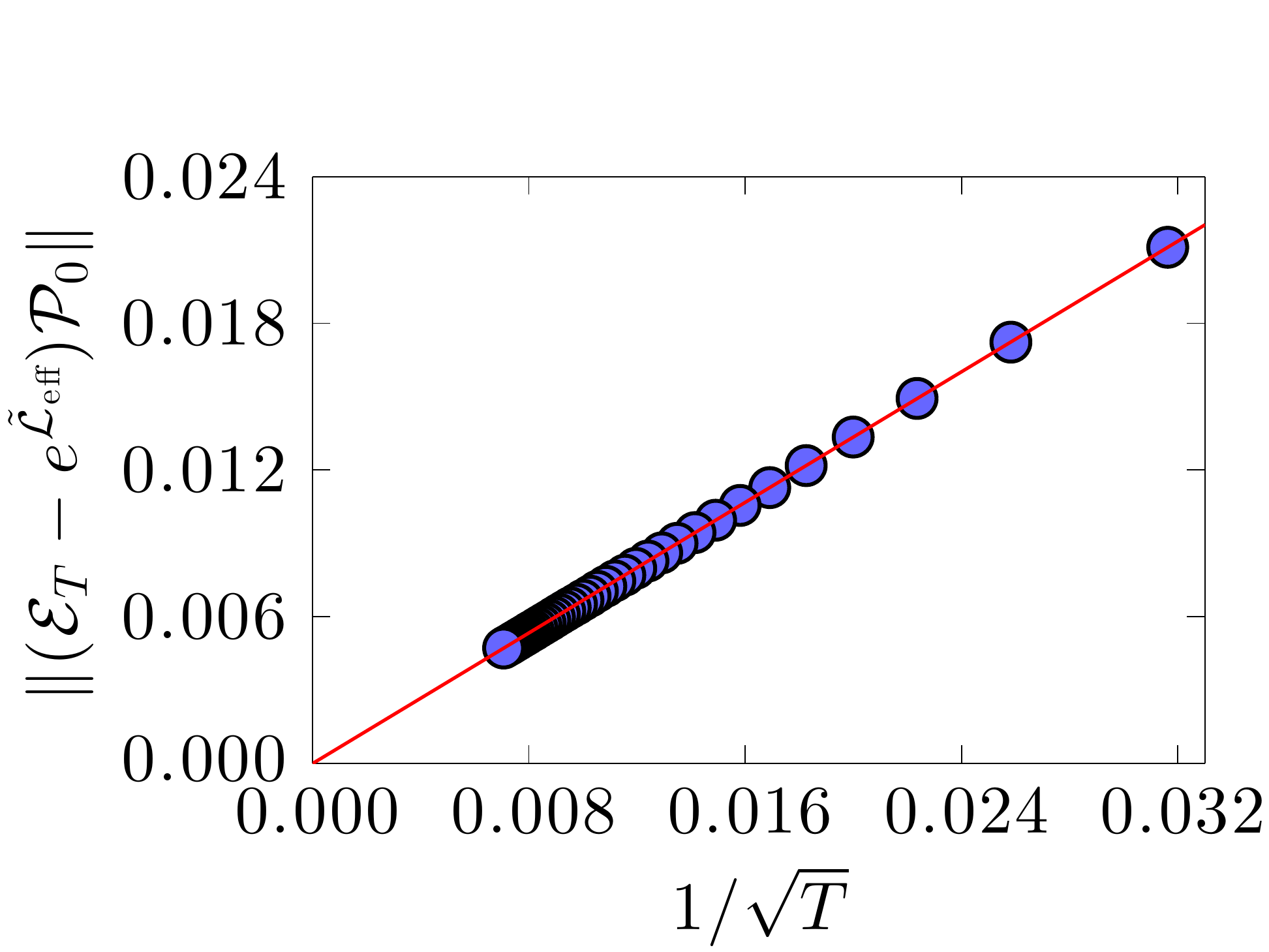}
\end{centering}
\vspace{-5mm}
\caption{(Color online) Distance from the exact evolution (${\cal E}_T := e^{T {\cal L}_T}$) and effective one with Liouvillian (\ref{thermal-2nd-order}) , as a function of $1/\sqrt{T}$. $N=3,\,\tau_+= 2,\,\tau_-=1$, and $g=(\tau_RT)^{-1/2}$ (where the relaxation time is $\tau_R=\frac{\tau_+ \tau_-}{\tau_+ + \tau_-})$. The linear fit is obtained using the least squares fitting on all of the data points, and the norm is the maximum singular value of the maps realized as matrices.}
\label{Fig:linear-thermal}
\end{figure}

{\em{Simulating Collective Amplitude Damping.--}} Here we use our general result Eq.~(\ref{L-simul}) to simulate a qubit subject to collective damping. This type of symmetric noise is interesting as it admits decoherence-free subspaces \cite{Zanardi:97c,Lidar:1998fk,Kielpinski:01} and can be used to dissipatively prepare entangled states.
Let us consider a system of $N$ qubits coupled to a bosonic mode
e.g., $N$ atoms coupled to a cavity EM mode, via a (collective) Jaynes-Cummings Hamiltonian
$K= g\, (S^- a^\dagger+S^+a)$
moreover we assume that the system dissipates according to the Liouvillian ${\cal L}_0=\mathbf{1}_S\otimes {\cal L}_B$ where
${\cal L}_B(\rho)=-i\omega\,[ a^\dagger a,\,\rho]+ \tau_R^{-1}(a\rho a^\dagger -\frac{1}{2}\{a^\dagger a,\,\rho\}).$
Using Eq.~(\ref{L-simul}) with $L_1=S^-$ one finds the effective generator
${{\cal L}}^{(S)}_{\textrm{eff}}(\rho)=4g^2\tau_R\,(S^-\rho S^+-\frac{1}{2}\{S^+S^-,\,\rho\})$
where $\rho$ is just the qubit state as the generator is trivial in the bosonic degrees of freedom (frozen at $|0\rangle.)$
\\
Proposition 2 shows that one can consider for the auxiliary qubits a Liouvillian that is more general than Eq.~(\ref{qubit-damping})
(as long as its steady state is unique). We illustrate this fact by considering the thermalization of an auxiliary qubit at non-zero temperature. Namely, 
 we add to Eq.~\eqref{qubit-damping} an excitation Liouvillian, such that now
${\cal L}_B(\rho)=\tau_{-}^{-1} (\sigma^-\rho\sigma^+ -\frac{1}{2}\{\sigma^+\sigma^-,\,\rho\})
+ \tau_{+}^{-1} (\sigma^+\rho\sigma^- -\frac{1}{2}\{\sigma^-\sigma^+,\,\rho\}).$
By explicit computation of the $\Gamma$ matrix in Prop. 2  one can check that
the new effective generator (in the system $S$ sector) is
\begin{equation}
{{\cal L}}^{(S)}_{\textrm{eff}}(\rho)=\sum_{\alpha=\pm} {\tau_{\textrm{eff},\alpha}^{-1}}( S^\alpha\rho S^{\alpha\dagger}-\frac{1}{2}\{S^{\alpha\dagger}S^{\alpha},\,\rho\}) 
\label{thermal-2nd-order}
\end{equation}
where $\tau_{\textrm{eff},\pm}^{-1} = 4g^2\frac{\tau_{-}\tau_{+}}{(\tau_{-}+\tau_{+})^2}\tau_{\mp}$. 
A numerical check of the validity of Eq.~(\ref{2nd-projection-th}) is shown in Fig.~\ref{Fig:linear-thermal}.  

Note that one can add a Hamiltonian term of the form $H_{B}=\omega_B \sigma^z$ in such a way that the unperturbed Lindbladian becomes a thermalizing Davies-type generator, for which the temperature is fixed by $\tau_{+}/\tau_{-} = \exp{(\beta_B \omega_B)}$ \cite{Alicki_book}.
In a similar fashion, using the remark {\bf{iii)}} above, we can add a properly rescaled Hamiltonian term ${\cal K}_1$ in such a way that $\tilde{\cal L}_{\textrm{eff}}\mapsto \tilde{\cal L}_{\textrm{eff}}' = \tilde{\cal L}_{\textrm{eff}}+ {\cal P}_0 \tilde{\cal K}_1{\cal P}_0$, and $\tilde{\cal L}_{\textrm{eff}}'$ is again of Davies type. In other words, also the effective generator can be made  thermal and so it defines a temperature according to $\frac{\tau_{\textrm{eff},+}}{\tau_{\textrm{eff},-}} =\exp(\beta_{\textrm{eff}} \omega_{\textrm{eff}})$ for some energy scale $\omega_{\textrm{eff}}$.  From $\frac{\tau_{\textrm{eff},+}}{\tau_{\textrm{eff},-}} = \frac{\tau_+}{\tau_-}$ it follows   $T_{\textrm{eff}}/T_B = \omega_{\textrm{eff}}/ \omega_B ,$ namely one has cooling or heating
according to whether $\omega_B > \omega_{\textrm{eff}}$ or vice-versa. 

%
%
%

{\em{Simulating collective dephasing.--}}
Following a similar set-up as the previous subsection but with a Hamiltonian of the form $K=g\, S^x \otimes \sigma^x$ ,
the effective  generator becomes that of collective dephasing along the $x$-direction
\begin{equation}
{{\cal L}}^{(S)}_{\textrm{eff}}(\rho)=4g^2\frac{\tau_- \tau_+}{\tau_+ +\tau_-}( S^x\rho S^x - \frac{1}{2}\{S^xS^x,\rho\}).
\label{bit-flip}
\end{equation}
%
%
%
\begin{figure}
\begin{centering}
\includegraphics[width=8cm]{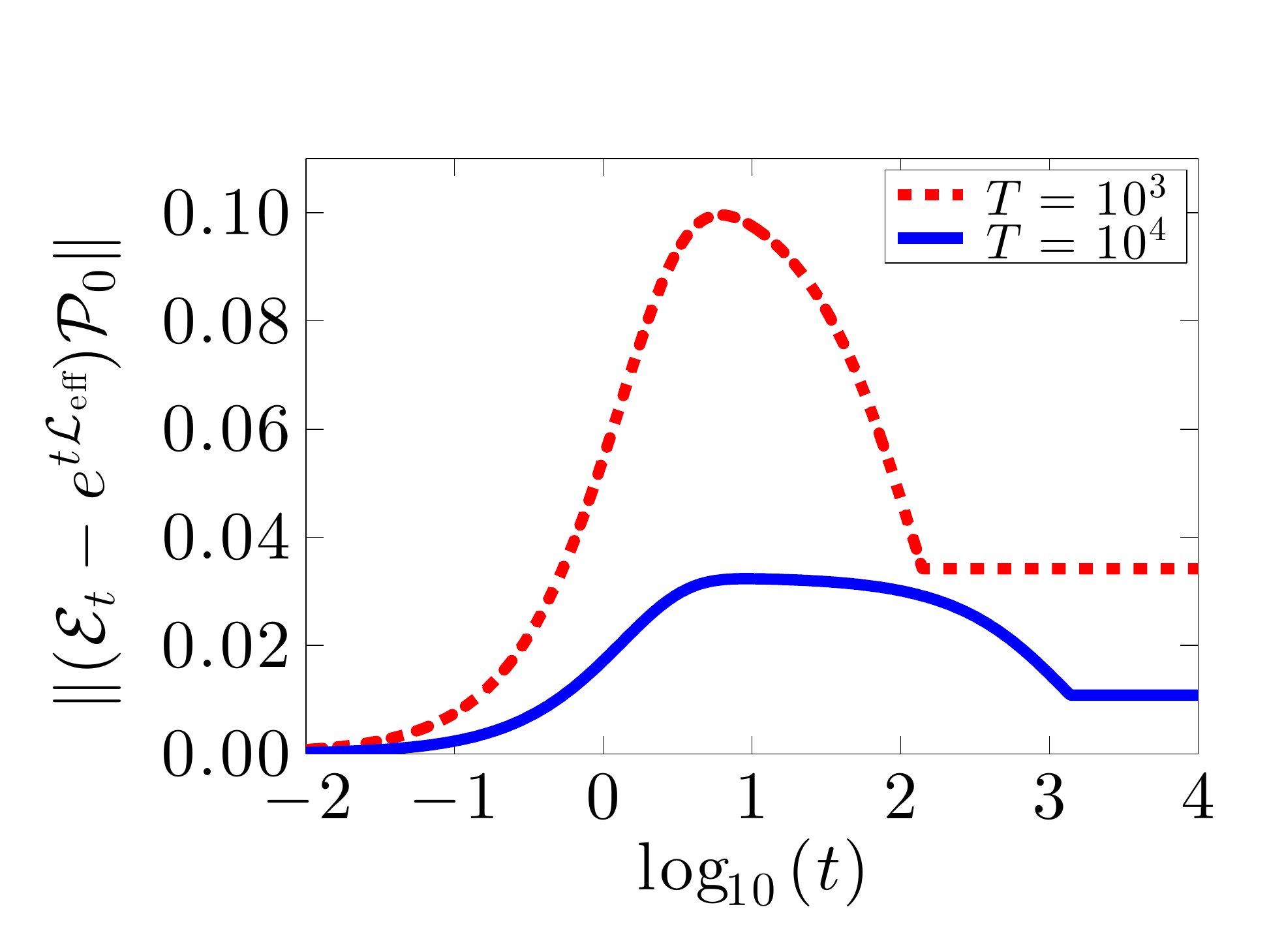}
\end{centering}
\vspace{-5mm}
\caption{(Color online) Distance from the exact evolution (${\cal E}_t := e^{t {\cal L}_T}$) and effective one with Liovillian Eq. (\ref{bit-flip}), as a function of $\log_{10}(t)$. $N=1,\,\tau_+=2,\,\tau_-=1$, and $g=(\tau_RT)^{-1/2}\, (\tau_R=\frac{\tau_+ \tau_-}{\tau_+ + \tau_-})$. Note that for the dashed line we have extended $t$ past $T$, purely for convenience. The norm is the maximum singular value of the maps realized as matrices.}
\label{Fig:eff-ideal}
\end{figure}

%

In Fig.~\ref{Fig:eff-ideal} we plot the distance between the actual and the effective evolution  as a function of $t$ for different time-scales $T$. 
According to Eq.~(\ref{2nd-projection-th}) by changing $T\rightarrow c\,T$ ($c>1$), we expect the distance to fall by a factor of $\sqrt{c}$ (cf.~in Fig.~\ref{Fig:eff-ideal} the maximum error falls from the dash to solid line by a factor of $\sim \sqrt{10}$).
In the limit of $T\rightarrow \infty$, the exact evolution becomes identical to that of the effective one for all times $t$ (the actual evolution `adiabatically follows' the effective one). 

%

{\em{Conclusions.--}}
 There is increasing evidence that dissipative and quantum incoherent processes can be used to enact quantum information processing primitives, see e.g. \cite{beige,Kraus-prep,kastoryano2011dissipative,barreiro2011open,verstraete2009quantum,dissi-top, Stannigel:2014rw,Yale}. 
In this paper we have shown how a suitable coherent coupling between a quantum system $S$ and an environment comprising multiple qubits subject to strong Markovian dissipation, can be used to simulate universal Lindbladian dynamics over $S.$ More precisely, by using high-order virtual dissipative processes,  one can build an effective Liouvillian generator in arbitrary Lindblad form \cite{Lindblad-paper} that  governs the dynamics of $S$ {\em{exactly}} in the limit of infinitely fast dissipation.
This approach has to be contrasted with the standard one in which Lindbladian evolution arises from a weak Hamiltonian coupling to a unitarily evolving environment
and involves uncontrolled Markovian approximations e.g., Born-Markov factorization of the joint density matrix \cite{Gardiner-Zoller}.
We illustrated our results by numerical simulations of concrete physical models.
Our findings show that Markovianity itself can be seen as resource in that it allows for universal simulation of an important class of quantum irreversible processes.

{\em{Acknowledgements.--}}
This work was partially supported by the ARO MURI Grant No. W911NF-11-1-0268.

{\em{Note Added. }} After the completion of this work we became aware of Ref.~\cite{sweke2015} where a different approach to simulation of Markovian systems is pursued.




\bibliographystyle{unsrt}



\begin{thebibliography}{10}
\bibitem{add-ham} From the Proof of Prop.~1 reported in the Appendix it easily follows
that only the term $T^{-1}{\cal P}_0 \tilde{\cal K}_1 {\cal P}_0$ is $O(1)$ whereas all the other
terms involving ${\cal K}_1$ are at most $O(1/\sqrt{T}).$
\end{thebibliography}

\newpage

\appendix
\section{Proof of Proposition 1}
Here we provide a Proof of Eq. (\ref{2nd-projection-th}) and an asymptotic (large $T$) estimate of the constant $C_\theta.$
Let $\cal P$ be the spectral projection of ${\cal L}_T={\cal L}_0+{\cal K},\,( \|{\cal K}\|=O(\frac{1}{\sqrt{T}} )$ associated with the zero eigenvalue of ${\cal L}_0$. 
Since, because of the Lindblad structure, there is no nilpotent term associated with the zero eigenvalue, the perturbation theory reads, as shown in In T.~Kato, {\em Perturbation theory for Linear operators}, for small $\|{\cal K}\|,$ i.e., large $T,$ 
\begin{eqnarray}
{\cal P}-{\cal P}_0&=& -{\cal P}_0 {\cal K} {\cal S}-{\cal S} {\cal K} {\cal P}_0+O(\|{\cal K}\|^2),\nonumber \\
{\cal P}{\cal L}_T{\cal P}&=&{\cal P}_0 {\cal K} {\cal P}_0-{\cal P}_0 {\cal K} {\cal S}{\cal K}{\cal P}_0 - {\cal P}_0 {\cal K} {\cal P}_0{\cal K}{\cal S}  \nonumber \\ & -& {\cal S} {\cal K}{\cal P}_0 {\cal K} {\cal P}_0 + O(\|{\cal K}\|^3) \label{Kato-Pert}
\end{eqnarray}
From the first equation it now follows (for sufficiently large $T$)
\begin{equation}
\|{\cal P}-{\cal P}_0\|=O(\tau_R\|{\cal K}\|)\le C_1^\prime\tau_R\|{\cal K}\| ,
\label{ineq0}
\end{equation}
where $ C_1^\prime$ is a suitable constant (notice that $\|{\cal P}_0\|=1$). On the other hand,
using ${\cal P}_0 {\cal K} {\cal P}_0=0$ and the definition   ${{\cal L}}_{{\textrm{eff}}}:= -{\cal P}_0 {\cal K} {\cal S}{\cal K}{\cal P}_0$ for the dimensionful effective generator
from the last equation in (\ref{Kato-Pert}) it follows 
\begin{equation}
\|{\cal P}{\cal L}_T{\cal P}-{\cal L}_{\textrm{eff}}\|=O(\|{\cal K}\|^3),
\label{ineq1}
\end{equation}
whence  (for small $\|{\cal K}\|$) $\|{\cal P}{\cal L}_T{\cal P}\|\le C_3 \| {\cal L}_{\textrm{eff}}\|.$
Since $e^{t {\cal L}_T} {\cal P}=e^{t {\cal P}{ \cal L}_T {\cal P} } {\cal P}$ one can write
\begin{align}
(e^{t {\cal L}_T} -e^{t {\cal P}{ \cal L}_T {\cal P} }){\cal P}_0 &=- (e^{t {\cal L}_T} -e^{t {\cal P}{ \cal L}_T {\cal P} })({\cal P}-{\cal P}_0)\nonumber\\
e^{t {\cal L}_T} -e^{t {\cal P}{ \cal L}_T {\cal P} } &=(e^{t {\cal L}_T} -e^{t{\cal L}_{\textrm{eff}} }) \nonumber\\
& +(e^{t{\cal L}_{\textrm{eff}} }-e^{t {\cal P}{ \cal L}_T {\cal P} })
\label{triangle1}
\end{align}
Using $\|e^{X}-e^{X+Y}\|\le\|Y\|e^{\|X\|+\|Y\|}$ with $X:=t {\cal L}_{\textrm{eff}}$ and $Y=t( {\cal P}{ \cal L}_T {\cal P}-   {\cal L}_{\textrm{eff}})$
and  the bounds above it also follows that, for $0\le t\le \theta T$, 
\begin{equation}
\|e^{t{\cal L}_{\textrm{eff}} }-e^{t {\cal P}{ \cal L}_T {\cal P} }\|\le C^\prime_2\, t  \,\|{\cal K}\|^3 
\label{ineq2}
\end{equation}
 where $C_2^\prime$ is a constant of 
that, for dimensional reasons, is $O(\tau_R^2)$ i.e., $C_2^\prime\le C_2 \tau_R^2.$  From (\ref{triangle1}) using $\|e^{t {\cal L}_T}\|=1,$ and standard operator norm inequalities
one finds  
\begin{widetext}
\begin{eqnarray}
\epsilon_t:= & &\| (e^{t {\cal L}_T} -e^{t{\cal L}_{\textrm{eff}} }){\cal P}_0\|\le \|(e^{t{\cal L}_{\textrm{eff}} }-e^{t {\cal P}{ \cal L}_T {\cal P} }){\cal P}_0 \|+ (\|e^{t {\cal L}_T}  \|+\|e^{t {\cal P}{ \cal L}_T {\cal P} }\|)\|{\cal P}-{\cal P}_0\|\nonumber \\ & & \le \|e^{t{\cal L}_{\textrm{eff}} }-e^{t {\cal P}{ \cal L}_T {\cal P} }\| +(1+e^{t \|{\cal P}{\cal L}_T{\cal P}\| })\|{\cal P}-{\cal P}_0\|
\end{eqnarray}
\end{widetext}
 Notice that $\epsilon_t$ is the quantity showing up in the LHS of (\ref{2nd-projection-th}) namely
is the quantity whose upper bound over $[0,\,T]$ we desire to show is $O(\sqrt{\tau_R/T}).$
Now using the bounds (\ref{ineq0}), (\ref{ineq1}),(\ref{ineq2}) and $0\le t\le \theta T, \,(\theta>0)$ one finds
\begin{equation}
\epsilon_t\le \tau_R \|{\cal K}\|( C_1+ C_2\,t\,\tau_R  \|{\cal K}\|^2),
\label{ineq3}
\end{equation}
where $C_1\ge C_1^{\prime}( 1+e^{ C_3 \theta \|\tilde{\cal L}_{\textrm{eff}}  \|}).$
 By moving to the dimensionless Hamiltonian such that $\|{\cal K}\|= (\tau_R T)^{-1/2} \|\tilde{\cal K}\|$
the inequality (\ref{ineq3}) becomes
$\epsilon_t\le \sqrt{ \frac{\tau_R}{T} }( C_1 + \frac{t}{T} C_2).
$
Notice that the requirement of $\|{\cal K}\|$ being sufficiently small used repeatedly in the above
translate now in the ``adiabatic criterion" of $T$ being sufficiently large.
Finally by taking the supremum for $t\in[0,\,\theta T]$ one obtains $\sup_t \epsilon_t\le  \sqrt{ \frac{\tau_R}{T} }\,(C_1 + C_2\,\theta).$
Setting $C_\theta:=C_1 +  C_2\,\theta $ completes the Proof of (\ref{2nd-projection-th}).


\section{Proof of Proposition 2}
We directly compute the second order effective generator ${{\cal L}}_{{\textrm{eff}}}:= -{\cal P}_{0}{{\cal K}}{\cal S}{{\cal K}}{\cal P}_{0}$  by acting on some state $X$, such that ${\cal P}_0 (X) = \rho \otimes {\rho}_0$. One has
${\cal S}{\cal K}{\cal P}_0(X) = -i \sum_{i=1}^M (L_i \rho \otimes {\cal S}_B (B_i \rho_0) - \\ \rho L_i \otimes {\cal S}_B (\rho_0 B_i)),$
where we have introduced notation ${\cal S} = \mathbf{1}_S \otimes {\cal S}_B$, which only acts non-trivially on system $B$ for this set-up, i.e. ${\cal S}_B = -\int_0 ^{\infty}e^{t{\cal L}_B} $ (following from ${\cal S} = -\int_0^{\infty} e^{t{\cal L}_0}{\cal Q}_0$, see \cite{zanardi-dissipation-2014}).
Acting with $-{\cal P}_0 {\cal K}$ on this we can see that:
\begin{widetext}
\begin{equation}
{\cal L}_{\textrm{eff}}(X) =  \{ \sum_{i,j = 1}^M \Gamma_{ij}^{(A)}\left(L_i \rho L_j - L_j L_i \rho\right) 
+ \sum_{i,j = 1}^M \Gamma_{ij}^{(B)}\left(L_j \rho L_i - \rho L_i L_j \right) \}\otimes \rho_0,
\label{2nd-order-general}
\end{equation}
\end{widetext}
where $\Gamma_{ij}^{(A)} = -{\mathrm{Tr}}\left( {\cal S}_B(B_i \rho_0)B_j\right)$, and $\Gamma_{ij}^{(B)} = -{\mathrm{Tr}}\left( {\cal S}_B(\rho_0 B_i)B_j\right)$.
In passing we notice that one can rewrite the system $S$ part of these equations in a more familiar form, using that without loss of generality $L_i=L_i^\dagger$ and $B_i=B_i^\dagger$. Just observe that since ${\cal S}_B$ is a Hermitian-preserving map, we have $\Gamma^{(A)*}=\Gamma^{(B)}$. Eq. (\ref{2nd-order-general-2}) then follows as required.

\section{ The dissipation-projection hierarchy}
Before concluding, we would like to show how the construction leading to the effective dynamics (\ref{2nd-projection-th}) can in principle be iterated
over a sequence of exponentially longer time-scales.  Let us set the error parameter $\epsilon:=\|K\|\tau_R\ll 1$.
The effective relaxation time of (\ref{2nd-projection-th})  can be roughly estimated as 
$\tau^{(1)}_R\sim \|{\cal L}_{\textrm{eff}}^{-1}\|\ge (\|K\|^2\tau_R)^{-1}=\epsilon^{-2}\tau_R\gg \tau_R
$,
where the last inequality stems from the condition $\epsilon\ll1.$ 

%
Suppose that  the dynamics generated by ${\cal L}_{{\mathrm{eff}}}$ admits itself a high-dimensional SSS (let 
${\cal P}_0^{(1)} $ denote the associated projection) and that   one can switch on an  {\em{extra}} Hamiltonian $K_1$ such that 
$\|K_1\|\tau_R^{(1)}=\epsilon\ll 1.$
One can now apply the projection theorem (\ref{2nd-projection-th}) to ${\cal L}_{\textrm{eff}}$ and $K_1$
and argue that the effective dynamics in the SSS of ${\cal L}_{\textrm{eff}}$ is ruled by ${\cal K}_{\textrm{eff}}^{(1)}:={\cal P}_0^{(1)} {\cal{K}}_1 {\cal P}_0^{(1)}.$
%
If even  this effective Hamiltonian vanishes then  one can {\em{iterate}} the projection procedure assuming as a starting state-space the  SSS of ${\cal L}_{\textrm{eff}}.$ In general, if at the $n$-th level one finds ${\cal P}_0^{(n)} {\cal K}_1 {\cal P}_0^{(n)}=0$ and $ {\cal P}_0^{(n)}$ is high-dimensional then one can  move to the next level where ${\cal L}_{\textrm{eff}}^{(n+1)} =- {\cal P}_0^{(n)}{\cal K}_n {\cal S}^{(n)} {\cal K}_n {\cal P}_0^{(n)}$ and a new Hamiltonian $K_{n+1}$ such that $\|K_{n+1}\|\tau_R^{(n)}=\epsilon\ll 1$ is introduced. Reasoning as in the above one can show that at each iteration the relaxation time scale (Hamiltonian norm) gets stretched (compressed) by a factor $\epsilon^{-2}$ ($\epsilon^{2}$).
%
%
From the point of view of potential applications the interest in exploring this projection hierarchy 
rests on the possibility that the first non-vanishing effective Hamiltonian has some desired property
e.g., higher-locality \cite{zanardi-dissipation-2014}. 

\end{document}